\documentclass[11pt]{article}
\usepackage{epsfig,amsmath,amssymb,graphics}
\newcommand{\tX}{\tilde{X}}
\newcommand{\half}{\frac{1}{2}}
\newcommand{\third}{\frac{1}{3}}
\newcommand{\GeV}{\text{ GeV}}
\newcommand{\preprint}[1]{\rule{0pt}{8pt} \scriptsize #1}

\begin{document}
\title{\bf The New Fat Higgs: Slimmer and More Attractive}
\author{Spencer Chang, Can Kilic, Rakhi Mahbubani\\
\small\sl Jefferson Physical Laboratory \\
\small\sl Harvard University \\
\small\sl Cambridge, MA 02138}
\date{}
\maketitle
\begin{picture}(0,0)
\put(400,200){\shortstack{
        \preprint HUTP-04/A025\\
        \preprint hep-ph/0405267\\
        \rule{0pt}{8pt} }}
\end{picture}

\begin{abstract}
In this paper we increase the MSSM tree level higgs mass bound to a value that is naturally larger than the LEP-II search constraint by
adding
to the superpotential a $\lambda S H_{u}H_{d}$ term, as in the NMSSM, and UV completing with new strong dynamics
{\it before} $\lambda$ becomes non-perturbative.  Unlike other models of this type the higgs fields remain
elementary, alleviating the supersymmetric fine-tuning problem while maintaining unification in a natural way.
\end{abstract}
\pagebreak

\section{Introduction}
Finding a satisfactory explanation for the large difference between the weak scale and the Planck scale, known as the
hierarchy problem, is an issue that has concerned particle physicists for more than two decades, and is the reason
why
the Standard Model higgs sector is widely held to be incomplete.  Supersymmetry (SUSY) provides arguably the most attractive solution for this hierarchy, since it comes with gauge coupling unification as an automatic consequence.  However its simplest implementation, the Minimal Supersymmetric Standard Model (MSSM), is looking increasingly
fine-tuned as recent results from LEP-II have pushed it to regions of parameter space where a light higgs seems unnatural.\footnote{See references
\cite{Giudice:2003nc,Barbieri:2003dd} for
further discussion.}  This is problematic for the MSSM since SUSY relates the quartic
coupling of the higgs to the electroweak gauge couplings, which at tree level bounds the mass of the lightest
higgs
to be less than that of the $Z$.  Radiative corrections can help increase this bound, with the largest contribution
coming from the top yukawa, giving
\begin{equation}\label{equ:quarticstop}
m^2_{h^0} \approx m^2_Z + \frac{3}{8\pi^2}h^4_t v^2\log{\frac{m_{\tilde{t}}^2}{m^2_t}}
\end{equation}
for large $\tan{\beta}$.
Since this effect is only logarithmic in the stop mass however, consistency with the LEP-II mass bound
requires the stops to be pushed up to at least 500 GeV.
At the same time radiative corrections to
$m_{H_u}^2$ are quadratic in the stop mass
\begin{equation}\label{equ:massstop}
\delta m^2_{H_u} \approx -\frac{3}{4\pi^2}m_{\tilde{t}}^2\log{\frac{\Lambda}{m_{\tilde{t}}}}
\end{equation}
There is therefore a conflict between our expectation that the stop is heavy enough to significantly increase the
higgs mass through radiative corrections and yet light enough to cut off the quadratic divergence
in a natural way.\footnote{A recent paper \cite{Birkedal:2004xi} attempted to resolve this conflict
by suppressing the size of radiative corrections to $m_{H_u}^2$ from the stop.}  Requiring consistency with LEP-II
results therefore forces us to live with a fine tuning of a few percent.

One way to resolve this issue is to generate a larger tree level quartic coupling for the higgses.  This can be accomplished through new F-terms as in the Next To Minimal Supersymmetric Standard Model (NMSSM)
\cite{Fayet:1974pd,Batra:2004vc}; new D-terms by charging the higgs under a new gauge symmetry \cite{Comelli:1992nu}; or by using ``hard'' SUSY breaking at low scales \cite{Polonsky:2000rs}.
We will choose to focus on the NMSSM, where the addition of a gauge singlet $S$ allows for the following
term in the superpotential
\begin{equation}
W=\lambda S H_u H_d
\label{eq:nmssm}
\end{equation}
and results in an additional quartic coupling for the higgses of the form $|\lambda|^2 |H_u H_d|^2$.  Unfortunately the requirement of perturbativity up to the GUT scale limits the size of $\lambda$ at the electroweak scale
\cite{Masip:1998jc} giving a maximum higgs mass bound of about 150 GeV.
This constraint was recently evaded in the Fat Higgs model \cite{Harnik:2003rs} by  allowing the coupling to become
nonperturbative at an energy lower than the GUT scale, where $S, H_u$ and $H_d$ were seen to be composites of new strong
dynamics.  All couplings were asymptotically free above this point and the higgs mass bound could be pushed up to 500 GeV.   On the other hand the composite nature of the higgs doublets gave rise to a different problem  -  gauge coupling unification was not
manifest and some ad hoc matter content had to be added to the theory to preserve it.  In addition, elementary higgs
fields needed to be reintroduced in order to generate the usual Standard Model yukawas at low energies.

In this paper, we will argue that UV completion of the NMSSM does not require us to sacrifice the desirable
properties of weak scale SUSY.  We will keep the higgs fields elementary, making unification manifest while
permitting the usual Standard Model yukawas to be written down.  Like the Fat Higgs, we use a composite $S$ but
instead we replace the $\lambda$ coupling above the compositeness scale by asymptotically free yukawas.  Since we
will no longer have to run $\lambda$,
which grows in the UV, all the way to the GUT scale, we can afford to start at a larger value at the electroweak
scale.  Unfortunately our scheme will require us to compromise slightly on how heavy we can make the higgs, but
this seems a small price to pay for natural gauge coupling unification.

The outline of the rest of the paper is as follows:  in Section
\ref{sec:model} we discuss the philosophy of this mechanism and
detail a specific model, in Section \ref{sec:discuss} we discuss the
bounds on the $\lambda$ coupling, and the issues  of fine tuning,
gauge coupling unification and the model's phenomenology. We
conclude in Section \ref{sec:conclusion}.

\section{Constructing a Model\label{sec:model}}
In SUSY models gauge contributions to anomalous dimensions are negative, tending to make yukawa couplings
asymptotically free.  The yukawas themselves, on the other hand contribute positive anomalous dimensions.
These competing effects, which are evident in the Renormalization Group Equation (RGE) for the NMSSM $\lambda$
coupling
\begin{equation}
\frac{d \lambda}{d t} = \frac{\lambda}{16 \pi^2}\left[4\lambda^2 + 3 h_t^2 - 3g_2^2 -\frac{3}{5}g_1^2
\right]+\cdots
\label{eq:nmssmrge}
\end{equation}
result in an asymptotically free $\lambda$ only when the gauge couplings involved are larger than $\lambda$
itself.  Even when they do not dominate the running, maximizing the negative contributions from the gauge sector
by adding as many $SU(5)$ $\mathbf{5} +\bar{\mathbf{5}}$ multiplets as are allowed by perturbative
unification gives an upper bound on the low energy $\lambda$ coupling \cite{Masip:1998jc}.  The benefit is
small here, however, since the electroweak gauge couplings remain quite weak for the majority of the running
and $g_3$ only affects $h_t$ at one loop.
This makes it difficult to significantly increase the low energy value of $\lambda$.

One way to improve the situation is to introduce new gauge dynamics through the following superpotential:
\begin{equation}
W_{\lambda} = \lambda_1\, \phi X H_u + \lambda_2\, \phi^c X^c H_d + M_X X X^c + M_{\tX} \tX \tX^c.
\end{equation}
We have added the fields $\phi, \phi^c,X,X^c,\tX,\tX^c$, which are charged under a new strong
gauge symmetry, with the $X$s also charged under the Standard Model as seen in Table \ref{prelimcharges}.
\begin{table}[t]
\centering
\begin{tabular}{l|c|c}
 & $SU(3)\times SU(2)_L\times U(1)_Y$ & $SU(n)_{s}$ \\
\hline
$\phi$ & $(1,1,0)$ & $\mathbf{n}$ \\
$\phi^c$ & $(1,1,0)$ & $\bar{\mathbf{n}}$ \\
$X$ & $(1,2,-\half)$ & $\bar{\mathbf{n}}$ \\
$X^c$ & $(1,2,\half)$ & $\mathbf{n}$ \\
$\tX$ & $(\bar{\mathbf{3}},1,\third)$ & $\bar{\mathbf{n}}$ \\
$\tX^c$ & $(\mathbf{3},1,-\third)$ & $\mathbf{n}$
\end{tabular}
\caption{Preliminary charge assignments for the new particles\label{prelimcharges}}
\end{table}
We choose $SU(n)$ to be our strong group as this permits our scheme to be most easily implemented.  Since the strong
gauge coupling ($g_s$) can now dominate the running, the $\lambda_{1,2}$ yukawas
can be asymptotically free for larger initial values and the resulting gain in $\lambda$ will be more substantial.
The two $X$ fields have been given a
supersymmetric mass, $M_X$, and are completed into
$(\mathbf{5},\mathbf{n})+(\bar{\mathbf{5}},\bar{\mathbf{n}})$
multiplets of $SU(5)\times SU(n)_s$ by the $\tX$s and thus maintain gauge coupling
unification.  Note that this doesn't require any MSSM particles to be
gauged under $SU(n)_s$.  The fields that are gauged under both the Standard Model
and the new group have large supersymmetric mass terms and thus
decouple from low energy physics.

Below the scale $M_X$ and $M_{\tilde{X}}$, integrating out the $X$s
and $\tX$s generates the nonrenormalizable operator %%
\begin{equation}
W_\text{eff} = -\frac{\lambda_1\lambda_2}{M_X} \; \phi \phi^c H_u H_d.
\label{eq:nonrenorm}
\end{equation}
There are two ways in which the NMSSM $\lambda$ coupling can be
obtained from this operator. One is to break $SU(n)_s$ by giving a
vev to $\phi$; as long as this breaking takes place close to the
$M_X$ scale, $\lambda$ can be satisfactorily large.  A simpler
approach, which we adopt in this paper, is to use the fact that
below $M_X,M_{\tX}$ there are 5 fewer flavors of the strong group,
making the gauge coupling get strong at low energies, forcing the
$\phi$ fields to confine into an NMSSM singlet which we will call
$S$.

Building a realistic theory from this philosophy is simply a matter of deciding what $n$ will be.
We use the fact that there is a restriction on the number of $SU(5)$ flavors that can be added to the
Standard Model for gauge couplings to perturbatively unify given that the added $SU(5)$ fundamentals do
not decouple until the TeV scale.\footnote{The
possibility
of a model with accelerated unification \cite{Arkani-Hamed:2001vr} and a lowered unification scale will not be
considered here.}
This requires 4 flavors or less and hence $n \leq 4$.  Another important constraint is on the number of flavors of
$SU(n)_s$ that remain after the 5 flavors in $X$ and $\tX$ have
been integrated out.  We want to avoid $n_f < n$ where
there is an Affleck-Dine-Seiberg vacuum instability \cite{Intriligator:1995au} and will ignore the potentially
interesting case $n_f = n$, where the Quantum Modified Moduli Space constraint might shed some light on
the $\mu$ problem.
Instead we
will
choose to start with $n+6$ flavors of $SU(n)_s$, where integrating out the 5 flavors gives $n_f = n+1$, making the
theory s-confine.  Now combining the requirement for asymptotic freedom
($n+6 < 3n$) with the perturbative unitarity constraint ($n \leq 4$) discussed earlier uniquely fixes $n=4$.\footnote{The
case of $SU(3)$ with 9 flavors might also be useful for our purpose.  This model
has been argued to have a linear family of conformal fixed points in
$(\lambda_i, g)$ space \cite{Leigh:1995ep} and would therefore be convenient when we discuss the possibility of
having a new superconformal fixed point in Section
\ref{sec:conformality}.  Alternatively if the $\tX$s required for unification were not also charged under the
strong group, satisfying the resulting constraints would be easier since we would have more room to maneuver.  However this theory is not naturally unified, and so will not be pursued here. \label{foot:strassler}}

\subsection{Details of the Model\label{sec:details}}
We now summarize the content and interactions of the model.  There is a strong $SU(4)_s$ gauge group, with the particle content shown in Table \ref{newparticles}.
\begin{table}[b]
\centering
\begin{tabular}{l|c|c}
& $SU(3)\times SU(2)_L\times U(1)_Y$ & $SU(4)_{s}$ \\
\hline
$\phi$  & $(1,1,0)$ & $\mathbf{4}$ \\
$\phi^c$ & $(1,1,0)$ & $\bar{\mathbf{4}}$ \\
$\psi_i$ for $i=1,\cdots,4$ & $(1,1,0)$ & $\mathbf{4}$ \\
$\psi_i^c$ for $i=1,\cdots,4$ & $(1,1,0)$ & $\bar{\mathbf{4}}$ \\
$X$ & $(1,2,-\half)$ & $\bar{\mathbf{4}}$ \\
$X^c$ & $(1,2,\half)$ & $\mathbf{4}$ \\
$\tX$ & $(\bar{\mathbf{3}},1,\third)$ & $\bar{\mathbf{4}}$ \\
$\tX^c$ & $(\mathbf{3},1,-\third)$ & $\mathbf{4}$ \\
\end{tabular}
\caption{Final charge assignments for new particles \label{newparticles}}
\end{table}
The superpotential contains
\begin{eqnarray}
W&=& W_{\lambda} + W_S + W_\text{d} \quad \text{where}\\
W_S &=& m\, \phi \phi^c \\ \nonumber
W_\text{d} &=&  y (T^i\, \phi \psi_i^c + T^{c\;i}\, \psi_i \phi^c + T^{ij}\, \psi_i \psi_j^c)+
   \\ && \frac{y'}{M_\text{GUT}^2}(\epsilon^{ijkl}\; T^B_i\, \phi \, \psi_j \psi_k \psi_l+
   \epsilon^{ijkl}\; T^{B^c}_i\, \phi^c \, \psi_j^c \psi_k^c \psi_l^c).
\end{eqnarray}
where we have introduced some singlets denoted by $T$.  After confinement,
$W_S$ gives a linear term in $S$ as in the Fat Higgs \cite{Harnik:2003rs} while
$W_\text{d}$ decouples the extra mesons by giving them mass terms with the singlets $T^i,T^{c\;i}, T^{ij}$.
Note that in the second line of $W_\text{d}$ there is a nonrenormalizable mass term for the baryons with the $T^B$s which is
suppressed by the GUT scale
$M_\text{GUT}$ and thus gives rise to light baryon states.  The constraints imposed by these light states will be
discussed in Section \ref{sec:phenomenology}.  Note that there is a non-anomalous $U(1)_R$ symmetry (under which
$\psi_i,\psi_i^c$ are neutral and all other $SU(4)_s$ flavors have charge 1) that
makes the given superpotential natural.

\subsection{Conformality and Confinement\label{sec:conformality}}

At high energies the strong group has 10 flavors and is within the conformal window
($\frac{3}{2} n < 10 < 3 n$) implying, in the absence of $\lambda_{1,2}$, that the theory flows to an interacting
fixed point in the IR \cite{Intriligator:1995au}.  As discussed previously the strong gauge coupling gives large
negative
contributions to the beta functions of the $\lambda_{1,2}$ couplings making them
asymptotically free for $g_s\gg\lambda_{1,2}$.  Ignoring electroweak couplings and the top yukawa, near Seiberg's
fixed point we have the RGE:
\begin{equation}
\frac{d \lambda_{1,2}}{d t} = \frac{7 \lambda_{1,2}^3}{16 \pi^2}
+ \gamma_* \lambda_{1,2} + \cdots
\label{eq:lamrge}
\end{equation}
The first term is the usual one loop term due to the yukawa
couplings while the second term contains contributions from all orders in the fixed point gauge
coupling $g_*$.  If the theory is at the fixed point then we have very precise information on the value of
$\gamma_*$ in the weak limit, since this is related to the $U(1)_R$ charges of the fields by the superconformal algebra.  Within the conformal window for example, $-1<
\gamma_* <0$ which indicates that the $\lambda_{1,2}$ couplings are relevant; they grow in the IR.
Our limited understanding of
strongly coupled theories prevents us from proceeding in full generality, so from now on we will restrict
ourselves to two plausible types of behavior.

The first possibility is the emergence of a new superconformal phase where both the new yukawas and gauge couplings hit fixed points in the IR; in this case it is hard to be quantitative about possible values of the NMSSM $\lambda$ coupling.
At best, we can specify a range of fixed point values of $\lambda_{1,2}$ which give
interesting $\lambda$ couplings,  without being able to justify if those values can be obtained.  Still, the
insensitivity of this scenario to UV initial conditions is very attractive.

In the second possibility the yukawa couplings get strong and
disrupt the conformality, pushing the theory away from the fixed
point.  In this case a reasonable bound on the sizes of
$\lambda_{1,2}$ can be given using their apparent fixed point values
from Eq. \ref{eq:lamrge}.  We will refer to this as the weak limit
bound. It is nontrivial that this bound on $\lambda$ will be large
enough to be of interest to us. In fact, the naive estimate will be
in the right range, but as we will see in Section
\ref{sec:higgsmassbound}, there are many unknown order one factors
that can change its size. An undesirable aspect of this case is that
the UV boundary conditions for $\lambda_{1,2}$ have to be tuned to
small values in order for these couplings to be just below their one
loop fixed points at low energies which saturates the weak limit
bound. This tuning could be improved somewhat if the gauge coupling
hits its fixed point at some intermediate scale. It is also worth
noting that this weak limit bound could give us a rough estimate of
the fixed point values of $\lambda_{1,2}$ in the first scenario.

At energies around the mass of the $X$s and their colored partners $\tX$, these 5 flavors
are integrated
out of the theory.
The terms in the RGEs for the supersymmetric masses typically give an ordering $m < M_X < M_{\tX}$.
Thus, the colored partners are integrated out first which leaves  7 flavors of $SU(4)_s$; this is still within
the conformal window and, in the electric description, has a stronger fixed point than the UV theory.  This would take
$|\gamma_*|$ from 1/5 to 5/7 and also increase the weak limit bound on $\lambda_{1,2}$
at the scale $M_X$.   For the coupling to approach this fixed point the ratio $M_X/M_{\tX}$ must be small.
As discussed in Section \ref{sec:unification}, there are no constraints on the size of this parameter from
unification as long as $M_X=M_{\tX}$ at the GUT scale.

Below $M_X$, $X$ and $X_c$ are integrated out and the theory becomes
a $SU(4)_s$ gauge theory with 5 flavors
$\Psi_I=(\phi,\psi_i),\Psi_I^c=(\phi^c,\psi^c_i)$ for $I = 0,\cdots,
4$. There is a dynamically generated superpotential %%
\begin{equation}
W_\text{dyn} = \frac{1}{\Lambda^7}\left[M_{IJ} B^I B^{c\;J} - \det M \right]
\end{equation}
written in terms of gauge invariant mesons ($M_{IJ} \sim \Psi_I \Psi_J^c$) and baryons
($B^I \sim \epsilon^{IJKLM} \Psi_{J} \cdots \Psi_{M}$).
At the scale $\Lambda \lesssim M_X$ this theory confines and the superpotential should be
written in terms of the canonically normalized meson and baryon fields.  Since the gauge coupling is strong the sizes
of the interactions after matching are in principle unknown.  However estimating their
sizes by Naive Dimensional Analysis (NDA)
\cite{Luty:1997fk} gives:
\begin{eqnarray}
W&=& W_\text{eff} + W_S + W_\text{d} +W_\text{dyn}\quad \quad \text{where}\\
W_\text{eff} &\to& \left[\sqrt{n}\; \frac{\lambda_1\lambda_2}{4\pi}\frac{\Lambda}{M_X}
\right] S H_u H_d \\
W_S &\to& \frac{m\Lambda}{4\pi} S \\
W_\text{d} &\to&  \frac{y \Lambda}{4\pi} (T^i M_{0i} + T^{c\;i} M_{i0} + T^{ij} M_{ij})+
   \frac{y'\Lambda^3}{4\pi M_\text{GUT}^2}(T^B_i B^i+T^{B^c}_i B^{c\;i}) \\
W_\text{dyn} &\to& \left[(4\pi) M_{IJ} B^I B^{c\;J} - \frac{(4\pi)^3}{\Lambda^2}\det M \right]
\end{eqnarray}
and we have defined $M_{00}$ to be $S$. The first two terms give us an NMSSM-like model at energy scales below
$\Lambda$.  In the $W_\text{eff}$ term we have done not only the normal NDA analysis, but also the large $n$ counting - notice that this partly compensates for the $4\pi$ NDA suppression.
Up to an unknown $O(1)$ constant, this results in a value for $\lambda$ at the confinement scale of:
\begin{equation}
\lambda = \sqrt{n}\; \frac{\lambda_1\lambda_2}{4\pi}\frac{\Lambda}{M_X}.
\label{eq:lambda}
\end{equation}
$W_S$ contains a term linear in $S$ that favors electroweak symmetry
breaking and explicitly breaks the Peccei-Quinn symmetry that would give rise to an undesirable light axion.
$W_\text{d}$ marries up the superfluous baryons and mesons with singlet $T$ partners as desired.
In addition, integrating out the heavy mesons will decouple their interactions in $W_\text{dyn}$.
It is also possible to add interactions that will give rise to the standard NMSSM $S^3$
coupling to eliminate the new $\mu$ problem arising from  the supersymmetric parameter $m$ but we will not
address this or the $\mu$ problems of $M_X$ and $M_{\tX}$ here.

\section{Discussion\label{sec:discuss}}
\subsection{$\lambda$ and the Higgs Mass Bound \label{sec:higgsmassbound}}
\begin{figure}
\begin{center}
\epsfig{file=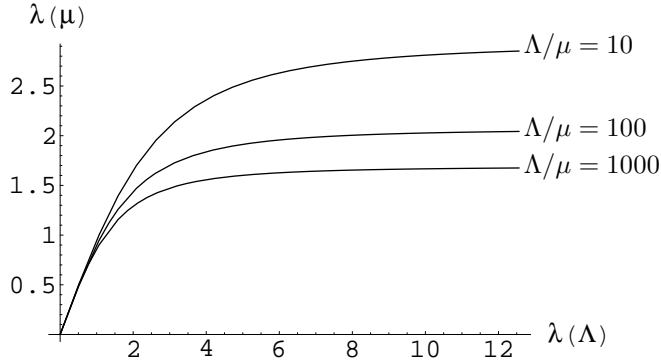} \caption{The low energy values of the
$\lambda$ coupling after running from the compositeness scale
$\Lambda$ down to the scale $\mu$.} \label{fig:lambda}
\end{center}
\end{figure}

So far we have shown how our model approximately reduces to the NMSSM below the
confinement scale.  Before analyzing this further it is important
to determine what range of $\lambda$ will be most useful for our purposes.
The value of the higgs quartic can be found by running the $\lambda$ coupling from the compositeness scale down to the electroweak scale ($\mu$).
We can solve for $\lambda$ in Eq. \ref{eq:nmssmrge} by ignoring all except the $\lambda^3$ term to obtain:
\begin{equation}
\lambda(\mu)^2 = \left(\frac{1}{\lambda(\Lambda)^2}+\frac{1}{2\pi^2}\ln{\frac{\Lambda}{\mu}}
\right)^{-1}.
\end{equation}
We summarize the resulting running in Figure \ref{fig:lambda}, in which the low energy value $\lambda
(\mu)$ is plotted as a function of the initial value $\lambda (\Lambda)$, for $\Lambda/\mu$ of different orders of
magnitude.  Notice that the value of $\lambda$ at low energies is largely insensitive to its value at the confinement
scale for $\lambda (\Lambda) \gtrsim 3$; it is this crucial feature that allows this model to compare favorably
with the Fat Higgs.
Unlike the Fat Higgs however, we do not have to start in the limit of strong coupling to get $\lambda(\mu)$
parameterically higher than the NMSSM bound of 0.8 \cite{Masip:1998jc}.  In the analysis that follows we will
arbitrarily choose as our region of interest $\lambda (\mu) \gtrsim 1.5$, which translates to
$\lambda (\Lambda) \gtrsim (1.8, 2.2, 3.3)$ for running over 1, 2, and 3 decades
respectively.
%%
%***********************************Lambda Figure**************************
%***********************************End of Figure**************************
%%

Returning to the first scenario in which there is a new superconformal fixed point, we can now relate the above
values of $\lambda$ to the fixed point values of $\lambda_{1,2}$.  Using Eq. \ref{eq:lambda} and assuming
comparable fixed points for the two yukawas, we see that we need
$\lambda_{1,2} \gtrsim (3.4, 3.7, 4.5)$ at $M_X$.
Unfortunately, we cannot say whether the actual fixed points satisfy this condition, although these values
are at least
feasible since the flatness of the RGE running of $\lambda(\mu)$ means that $\lambda$ does not have to
equal $4\pi$ at the confinement scale.
It would be interesting to do a more detailed study to determine whether this occurs.

It is possible to be more quantitative than this in the second case by relying on
our knowledge of the model in the weak limit.  Using Eq. \ref{eq:lamrge} we see that
\begin{equation}
\lambda (\Lambda) \sim \sqrt{4}\; \frac{\lambda_1\lambda_2}{4\pi}\frac{\Lambda}{M_X} \lesssim
 -\frac{\Lambda}{2\pi M_X} \frac{16\pi^2}{7}\; \gamma_* \sim \frac{8\pi}{7} \gamma_* \sim 3.6 \; \gamma_*.
\end{equation}
If we start with all 10 flavors of $SU(4)_s$ we have $\gamma_* = -1/5$ and $\lambda(\Lambda) \lesssim 0.7$ which is
too low to be of interest.  However, integrating out the $\tX$s leaves us with 7 flavors, which at the fixed
point gives $\gamma_* =
-5/7$ and $\lambda(\Lambda) \lesssim 2.6$.
We saw that this gives rise to a $\lambda$ that is in the interesting range for almost 3 decades of running between the confinement scale and the electroweak scale, suggesting that there are regions of parameter space where the low energy $\lambda$
coupling is large enough to be of interest.

We can calculate the tree level bound on the higgs mass by assuming that we are somewhere in the region  $1.5
\lesssim \lambda(\mu) \lesssim 2$ and using the NMSSM equation:
\begin{equation}
m^2_h \leq m_Z^2 \cos^2{2\beta} + \lambda^2 v^2 \sin^2{2\beta}/2.
\end{equation}
to obtain
\begin{equation}
m_h \lesssim 260-350 \text{ GeV}
\end{equation}
which is a substantial improvement over the MSSM bound of 90 GeV.
Taking the largest $\lambda(\mu)$ in Figure \ref{fig:lambda} pushes this bound up to 490 GeV, but
this is probably less generic in the parameter space.  Radiative corrections from the top sector
can increase this further although these are no longer necessary to satisfy the LEP-II bound.

We emphasize that it is rather surprising to obtain interesting results in the weak limit bound in spite of the NDA suppression factor of $4 \pi$.  This is a direct consequence of $\lambda$ not having to start off at $4\pi$; moderately large coupling is sufficient.  However, the robustness of our conclusions in the weak limit depends on a number of $O(1)$ unknowns which we ignored in the above analysis.  These are listed below and discussed in turn.
\begin{itemize}
\item {the value of the factor $\Lambda/ M_X$}
\item {the running of the nonrenormalizable operator in Eq. \ref{eq:nonrenorm} due to gauge coupling contributions in the region  $\Lambda\leq E\leq M_X$}
\item {the coefficient in the NDA matching that was used in Eq. \ref{eq:lambda}}
\item {loop-level corrections from $g_*$ to the coefficient of $\lambda_{1,2}^3$ in Eq. \ref{eq:lamrge}.}
\item {restrictions due to the large top yukawa.}
\end{itemize}
The first tends to suppress the value of $\lambda$ at low energies.
The strong dynamics after flavor decoupling suggests that this
factor is close to one, but it cannot be determined exactly since we
do not have detailed information on the fixed point value and exact
running of $g_s$ below $M_X$. It might, however, be compensated by
the effect of the second which enhances $\lambda$, hence we might be
able to make a case for neglecting them both, especially since this
allows us to make a quantitative prediction.  The $O(1)$ coefficient
in the third item parametrizes our ignorance of the physics of
strong coupling and unfortunately cannot be eliminated.  The fourth
point is that we ignored higher order gauge corrections to the
$\lambda_{1,2}^3$ term in Eq. \ref{eq:lamrge} at the gauge coupling
fixed point.  If the coefficient of this term decreases, the upper
bound on the $\lambda$ coupling increases and vice versa. Notice
however, that higher loop $\lambda_{1,2}$ corrections to the RGE are
suppressed and have been rightfully ignored since the loop
suppression factor $\lambda_{1,2}^2/(16\pi^2) \lesssim -\gamma_*/7
\leq 1/7 \ll 1$.  Finally, the fact that the top yukawa is not
neglible at low energies places some constraints on how large we can
make $\lambda_1$ without losing perturbativity for both these
couplings to the GUT scale. Doing a simple one loop analysis, for
$\tan \beta$ near 1 (where the gain in the tree level bound is
greatest), the $\lambda_1$ fixed point is about half of the value in
the above analysis which in turn halves the size of
$\lambda(\Lambda)$.  In general, we expect that there is some $O(1)$
suppression from this effect, but there is no comparable suppression
in $\lambda_2$ due to the smallness of the bottom yukawa.   Although
it is unfortunate that these factors cannot be evaluated to
determine a more specific bound, that the naive answer is in the
interesting range suggests that the actual value of $\lambda$ can be
similarly large.

Since we were motivated to explore this model by concerns of naturalness, we will now discuss how this
scenario helps the fine tuning.
First of all, the higgs mass bound has increased so it is no longer necessary for the
top squarks to be made heavy to evade the LEP-II bound.  In fact, it is now possible for all the MSSM scalars
including the higgs to have masses that are of the same order.  Thus, from a bottom-up perspective, there
are no unnatural hierarchies in these masses.\footnote{It could be argued that the top-down approach is still
problematic since starting with universal scalar and gaugino masses ($m_0$ and $m_\half$)
at the unification scale,  for example, force the top squarks to be heavy given observational lower bounds on
chargino and slepton masses.
This is a property of current SUSY breaking scenarios,
however, and it is possible to imagine alternatives with more random boundary conditions at the GUT scale that
result in realistic particle spectra with light top squarks.}
On the other hand there is a new fine tuning introduced in the weak limit (the second scenario), since
the UV initial conditions have to be precisely tuned to avoid breaking conformality.  However, these
parameters are at least technically natural and so could still have the right size.  There is no such fine
tuning in the
new
superconformal phase since the attractive IR fixed points reduce the sensitivity to UV
initial conditions.  For further discussion of how a larger higgs quartic coupling helps the fine tuning
issue see \cite{Bastero-Gil:2000bw} and Casas et. al. in \cite{Polonsky:2000rs}.

\subsection{Gauge Coupling Unification\label{sec:unification}}
In both the Fat Higgs and the New Fat Higgs SUSY guarantees that running the SM gauge couplings through the
strong coupling regions does not  give corrections larger than typical threshold effects.  We will recount the
argument here for
completeness.
Matching holomorphic couplings of a high energy theory containing a massive field with those of a low energy
theory with the field integrated out, is constrained by holomorphy.
In
particular, the matching depends only on the bare mass of the field and thus is not affected by strong
dynamics
\cite{Novikov:uc}.
For instance, taking $M_X = M_{\tX} = M$ at the cutoff $M_\text{GUT}$,
the high and low energy SM gauge couplings
(with and without the $X,\tX$ respectively) are matched at the bare mass $M$:
\begin{eqnarray}
g_{\text{sm,\,le}}(M) = g_{\text{sm,\,he}}(M)
\label{eq:matching}
\end{eqnarray}
where the high energy gauge couplings have their unified value at $M_\text{GUT}$.
At other energies these holomorphic couplings are determined by their one loop running
(with beta functions $b_\text{i,\,le} = b_\text{i,\,MSSM}$ and $b_\text{i,\,he} = b_\text{i,\,le}
+4$).
However,
during this running
the coefficients of the matter kinetic terms ($Z$) can change.
Thus to reach a more ``physical'' coupling, one should go to canonical normalization for the matter fields.
This rescaling is anomalous and relates the couplings by
\begin{eqnarray}
\frac{8\pi^2}{g_\text{le,\,phys}^2} = \frac{8\pi^2}{g_\text{le}^2} - \sum_i T^i \ln Z_i
\end{eqnarray}
where $i$ only runs over the matter fields in
the low energy theory and the $T_i$s are their Dynkin indices.
All potential strong coupling effects are contained within the $Z_i$s of the low energy
fields.  As a matter of fact,
there is actually no effect due to the RGE splitting $M_X < M_{\tX}$, since the matching in Eq.
\ref{eq:matching} of the low energy couplings occurs at $M$, giving no restriction on the ratio of these masses from unification.
An order one  $\ln Z_i$ gives a contribution of the order of a typical theshold correction; thus it takes
exponentially large $Z_i$ to adversely affect unification.  In this model,
such large $Z_i$ can only occur for the higgses when the $\lambda_{1,2}$ couplings are
strong for an exponentially large region.   Thus, the weak limit case is generically safe whereas in the new
superconformal phase, the conformal region for $\lambda_{1,2}$ cannot be exponentially large without affecting
unification.  Note that a similar constraint applies to the conformal region in the Fat Higgs model.

Aside from this potential constraint, gauge coupling unification occurs naturally
in this theory since the additional matter is charged under
the SM in complete $SU(5)$ multiplets {\it and} because the higgses are elementary (hence the beta functions of the
SM couplings are equivalent to those of the MSSM up to $SU(5)$ symmetric terms as detailed earlier).
In comparison, the Fat Higgs model had elementary preons which correctly reproduced the
running of the higgs doublets above the compositeness scale, but also contained additional fields which were put into both split
GUT and non-GUT multiplets in order to restore unification.
In that model, explaining why unification is natural requires a setup that generates the additional
matter content as well as
the required
mass spectrum.

\subsection{Phenomenology \label{sec:phenomenology}}
Much of the phenomenology in this model is similar to the Fat Higgs.  In both theories the
physics at the TeV scale is NMSSM-like with a linear term in $S$ but no cubic.  The
low energy $\lambda$ coupling is large and gets strong before the GUT scale, but
some asymptotically free dynamics takes over to UV complete the theory.  They both have similar higgs spectra which
are in concordance with precision electroweak constraints.
Also,  the analysis in  \cite{Kitano:2004zd} which concludes that UV insensitive
Anomaly Mediation works in the Fat Higgs should also apply to this model.

One notable difference between the two models is the additional baryon physics in our model.  The $B^0$ and $B^{c\;0}$
in this theory  get a large supersymmetric mass
from the $S$ vev and are not problematic.
However we also have light baryon states, the four $B^i$s and $B^{c\; i}$s that are married to the
$T^{B}_i$s and $T^{B^c}_i$s,
with supersymmetric masses of order
\begin{equation}
M_B \sim \frac{\Lambda^3}{4\pi M_\text{GUT}^2} \sim 10^{-13} - 10^{-7} \text{ eV}
\end{equation}
for $\Lambda \sim 5 - 500 \text{ TeV}$.  The scalar components of
these chiral superfields get TeV sized soft masses from SUSY
breaking and it is possible to determine these from the masses of
the elementary fields using the the techniques in
\cite{Arkani-Hamed:1998wc}.  The fermionic components are more
worrying since they remain light and thus give rise to some
stringent cosmological constraints. For instance, they decouple at a
$T_{\text{dec}} \sim 10 \GeV$, requiring $T_\text{reheat} \lesssim
T_\text{dec}$  in order to be consistent with Big Bang
Nucleosynthesis. It is also unclear whether the LSP in this theory
is a good Dark Matter candidate, given that it is never produced
with thermal abundance, or whether baryogenesis can be made to work
given such a low reheat temperature.

This constraint on the reheat temperature can be relaxed by adding
small mass terms for the fundamental fields of the form $W\supset
m_{IJ}\Psi_I \Psi_J^c$ which become tadpoles for the mesons after
confinement. The tadpoles induce meson vevs which give masses to the
light baryonic states through $W_\text{dyn}$.\footnote{We thank
Manuel Drees for proposing this solution to us.} These mass terms
break the $U(1)_R$ symmetry mentioned in Section \ref{sec:details},
however even very small masses ($m_{IJ}\sim M_B\gtrsim 1 MeV$)
ensure that Big Bang Nucleosynthesis can proceed as normal, while
the newly added masses are small enough for the symmetry breaking
effects to be under control.

It is also possible to circumvent this issue by using the scenario
with the Quantum Modified Moduli Space mentioned in Section
\ref{sec:model} or by making models without baryons, for instance
with an $Sp(2) \equiv SO(5)$ theory, starting with 18 fundamentals
of the $Sp(2)$. Integrating out the $X,\tX$ will reduce to the
s-confining case with 8 fundamentals.  At high energies, this has a
vanishing one loop beta function but is not asymptotically free at
two loops. With all 18 fundamentals and their yukawas, the analysis
in \cite{Leigh:1995ep} suggests that there is a superconformal fixed
point for the yukawa and gauge couplings. Specifically there is a
linear family of fixed points which run through the free fixed point
$(g = 0, \lambda_i = 0)$ (see Footnote \ref{foot:strassler}) and it
needs to be determined whether the fixed point values of
$\lambda_{1,2}$ are large enough to be in the interesting range.  We
can also work in a limit analogous to our weak limit of the previous
section, integrating out the $\tX$s first; this leaves the group in
the conformal window with 12 fundamentals.  Thus if $M_X/M_{\tX}$ is
small enough the theory can run to Seiberg's strong conformal fixed
point before the $X$s are integrated out.  In this case, the weak
limit bound gives $\lambda(\Lambda) \lesssim 1.8$, so we would need
$M_X$ near the weak scale or some help from the unknown order one
contributions detailed above.  However, since there are no baryons
in $Sp(n)$ theories we only have to to decouple the extra mesons.
From this reasoning we see that the physics associated with the
baryons does not appear generic to all implementations of our
mechanism and thus cannot be used to rule out all models of this
type.

\section{Conclusion\label{sec:conclusion}}
\begin{figure}
\begin{center}
\epsfig{file=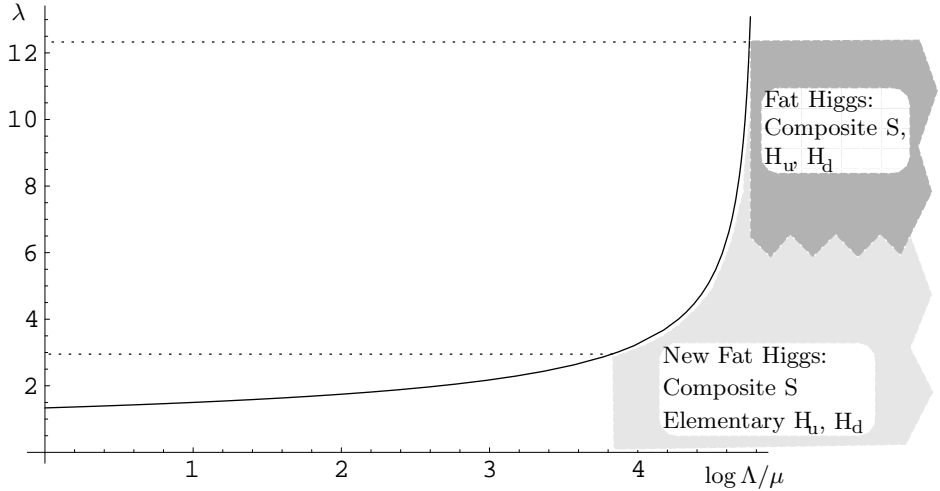} \caption{A comparison of UV completion scales
in the Fat Higgs and the New Fat Higgs} \label{fig:lambdacomp}
\end{center}
\end{figure}

Supersymmetry does extremely well in solving the hierarchy problem, but as
more precise measurements have told us, the minimal implementation of weak scale supersymmetry (the MSSM)
is becoming fine tuned at about the percent level.  Approaches that attempt to alleviate this problem have been many and varied, all of which have their own advantages and disadvantages.  Led by the
positive aspects of the MSSM, we analyzed a UV
complete NMSSM model which justifies the presence of a large $\lambda$ at low energies, resulting in a similarly
large higgs quartic coupling.  We did this by splitting the $\lambda$ coupling
into two asymptotically free yukawa couplings, allowing the theory to be continued above the apparent
strong coupling scale.  The simple model pursued in this paper is similar in
spirit to the Fat Higgs model: we start at the electroweak scale with a large $\lambda$ coupling which grows with increasing energy scale.  Rather than waiting for it to hit $4\pi$ before UV completing, we do this at a lower scale, leaving a theory with a composite S only (see Figure \ref{fig:lambdacomp}).
%***********************************Lambda Figure**************************
%***********************************End of Figure**************************
There is no need for a dynamically generated
superpotential because the induced $\lambda$ coupling never becomes non-perturbative; instead moderately strong coupling is sufficient to achieve a large tree level higgs mass bound without making the higgs fields composite.  This results in a higgs that is not as
fat as in the Fat Higgs, but gauge coupling unification, arguably the best evidence for weak scale SUSY,
is naturally maintained.

We did not study in depth the potentially interesting scenario where
the theory hit a superconformal fixed point, since it was tricky to
make any definitive statements about the fixed point values of
$\lambda_{1,2}.$ The strong coupling dynamics also made it difficult
to give exact results in the second case we considered, but we were
able to set a reasonable upper bound on $\lambda$ at low energies,
up to some unknown order one coefficients,  using the properties of
Seiberg's fixed point and superconformality in the weak limit. That
this bound turned out to give large enough $\lambda$ is comforting,
since it suggests the possibility of realizing our mechanism for a
generic parameter space with similar results.  However, to say any
more requires a detailed understanding of both the RGE equations at
strong coupling and matching at the confinement scale.

Finally, we discussed some of the implications of our model.  We saw
that the fine tuning issue was indeed ameliorated, at least from a
bottom-up perspective and that unification was not affected by the
strong coupling.  We also discussed the equivalence of the
phenomenology to that of the Fat Higgs Model in that there was
little difference in their higgs spectra or compatibility with
precision electroweak constraints.  One notable difference was the
presence of light fermionic baryons in our theory.  It would be
interesting to analyze the new baryon physics in more detail,
especially since they give rise to an interesting cosmological
constraint. As discussed, this can be relaxed by adding mass terms
that weakly break the non-anomalous $U(1)_R$. Furthermore, the
existence of models which do not have baryons suggest that light
states are not generic to this framework. In such models we expect
the dark matter abundance and baryogenesis analysis to proceed along
the lines of \cite{Menon:2004wv}.

In a few years the LHC will start to explore the possible presence of weak-scale supersymmetry
and it is important to continue to study SUSY models so as to
compare data with experiment.
Using naturalness as a guideline, it already seems that the simplest SUSY models are fine-tuned,
which motivates us to attempt to generalize them.  With this intuition we have analyzed a theory which improves the
naturalness of weak scale SUSY in a simple way without losing the natural unification of the MSSM.  However, only
experiment can ultimately determine the accuracy of our guesses for what comes beyond the Standard Model.

{\it Note:  As this paper was being finished, a paper appeared that analyzed a very
similar model \cite{Kobayashi:2004pu}.  However, they did not
notice the mechanism we have described for generating the NMSSM $\lambda$ coupling and had too
few flavors of $SU(4)_s$ to avoid the Affleck-Dine-Seiberg vacuum instability after
integrating out the $X$ fields.  Also, their
results on the suppression of the $\mu$ and $m_{H_u}$ corrections are not as crucial when the tree level upper bound on the higgs mass is increased.}

\section*{Acknowledgements}
We thank Zackaria Chacko, Markus Luty and Matt Schwartz for many
useful discussions. We would also like to thank Hitoshi Murayama for
his talk on the Fat Higgs at Harvard which inspired this project.
Most importantly, we want to thank Nima Arkani-Hamed for many
enlightening discussions and comments, all of which he provided in
spite of his recent penchant for committing unnatural acts
\cite{Arkani-Hamed:2004fb}. SC's work is supported by a NSF Graduate
Student Fellowship.

\end{document}